\documentclass{elsart}
\usepackage{graphicx,amssymb}
\begin{document}
\begin{frontmatter}

\title{Consensus formation on a triad scale-free network}
\author{A.O. Sousa\thanksref{label}}
\thanks[label]{{\it E-mail address:} sousa@ica1.uni-stuttgart.de}

\address{Institute for Computer Physics (ICP), University of 
Stuttgart, \\ Pfaffenwaldring 27, 70569 Stuttgart, Germany.}

\begin{abstract}
Several cases of the Sznajd model of socio-physics, that only a group of 
people sharing the same opinion can convince their neighbors, have been 
simulated on a more realistic network with a stronger clustering. In 
addition, many opinions, instead of usually only two, and a convincing 
probability have been also considered. Finally, with minor changes we 
obtain a vote distribution in good agreement with reality. 
\end{abstract}

\end{frontmatter}

\section{Introduction}
A growing interest has focused on the statistical physics of complex networks. 
They describe a wide range of systems in nature and society, modeling diverse 
systems as the world wide web, the net of human sexual contacts, or a network 
of chemicals linked by chemical reactions \cite{barabasi}. Such networks 
posse a rich set of scaling properties. A number of them are scale-free, that 
is, the probability that a randomly selected node has exactly $k$ links decay 
as a power law, following $P(k)\sim k^{-\gamma}$, where $\gamma$ is the degree
exponent, in consequence they present resilience against random breakdowns,
this effect is known as robustness of the network. The shortest-path-length
between their sites grows slowly (i.e. logarithmically) with the size of the 
network. That is, in spite of large sizes of networks, the distance between 
their sites are short - a feature known as the ``small world'' effect.

\begin{figure}[!h]
\begin{center}
\includegraphics[angle=-90,scale=0.55]{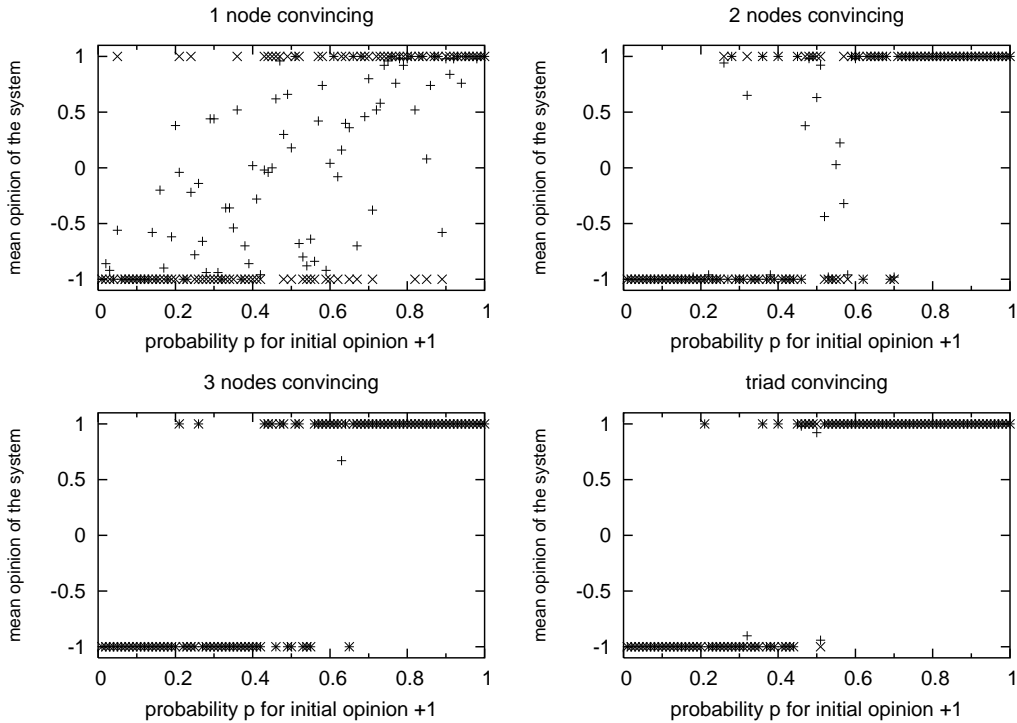}
\end{center}
\caption{Mean opinion of the system for 1 node (on top left), 
2 nodes (on top right), 3 nodes (on bottom left) and triad (on bottom right) 
convincing, and different initial concentrations $p$ of opinion $+1$ after 
$t=1$ (+) and $t=10$ (x) timesteps.}
\label{fig:rule_general}
\end{figure}

The majority of networks used to generate a scale-free topology are 
stochastic, they create networks in which the nodes appear to be randomly 
connected to each other. Scale-free random networks have naturally a 
continuous degree distribution spectrum, but recently it has been shown 
that discrete degree distributions of some deterministic graphs also have 
a power law decay \cite{barabasi}. Furthermore, scale-free random 
networks are excellently modeled by such deterministic graphs 
\cite{Doro:cond-matt}. However the comparison between the behavior of 
stochastic and deterministic networks in the simulation of a particular 
model still remains open. 

The involvement of physics in research on spreading of opinion's models 
has been a matter of increasing studies nowadays (see \cite{stauffer} 
for a review). Of particular interest here is the Sznajd model 
\cite{sznajd1}, which is one of several recent consensus-finding models 
\cite{defuant} and in which each randomly selected pair of nearest 
neighbors convinces all its neighbors of the pair opinion, only if the 
pair shares the same opinion; otherwise, the neighbor opinions are not 
affected. One time step means that on average every lattice node is 
selected once as the first member of the pair. It differs from other 
consensus models by dealing only with communication between neighbors, 
and the information flows outward, as in rumors spreading. In contrast, 
in most other models the information flows inward. Initially, two 
opinions ($\pm 1$) are randomly distributed with probability $p$ over 
all the nodes of the lattice. The basic Sznajd model with random 
sequential updating always leads to a consensus (all sites have opinion 
$+1$ or $-1$ and the whole system reaches a fixed point after a certain 
time of simulation). A phase transition is often observed as a function 
of the initial concentration of opinion $p$. A generalization to many 
different opinions (instead of only $\pm 1$) put into a Barab\'{a}si-Albert 
network \cite{barabasi} reproduced quite well the results of the
complex elections of city councillors in the state of Minas Gerais in 
Brazil \cite{bernardes}.

\section{The model}
 
\subsection{The triad network}

Although the Barab\'asi-Albert \cite{barabasi} network has successfully
explained the scale-free nature of many networks, a striking
discrepancy between it and real networks is that the value of the
clustering coefficient - which is the probability that two nearest 
neighbors of the same node are also mutual neighbors - predicted by the
theoretical model decays very fast with the network size and for large 
systems is typically several orders of magnitude lower than found 
empirically (it vanishes in the thermodynamic limit
\cite{kertesz,stauffer3}). In social networks, for instance, the clustering 
coefficient distribution $C(k)$ exhibits a power-law behavior, 
$C(k) \propto k^{-\gamma}$, where $k$ is number of neighbors (degree or 
connectivity) of the node and $\gamma \approx 1$ (everyone in the network 
knows each other). 

Very recently, by adding a triad formation step on the Barab\'asi-Albert 
prescription, this problem has been surmounted and scale-free models with 
high clustering coefficient have been investigated \cite{kertesz}. The 
Barab\'asi-Albert network starts with a small number $m$ ($m=3$ in our 
simulations) of sites (agents, people) all connected with each other. 
(For $m=2$, $4$ and $5$ we observed similar results for 100 runs
each.) Then a large number $N$ of additional sites is added 
as follows: First, each new node (node $i$) performs a preferential 
attachment step, i.e, it is attached randomly to one of the existing
nodes (node $j$) with a probability proportional to its degree; then 
follows a triad formation step with a probability $p_{\rm tf}$: the
new node $i$ selects at random a node in the neighborhood of the one
linked to in the previous preferential attachment step (node $j$). If 
all neighbors of $j$ are already connected to $i$, then a preferential 
attachment step is performed (``friends of friends get friends''). In this 
model, the original Barab\'asi-Albert network corresponds to the case of 
$p_{\rm tf}=0$. It is expected that a nonzero $p_{\rm tf}$ gives a 
finite nonzero clustering coefficient as $N$ is increased, while the 
clustering coefficient goes to zero when $p_{\rm tf}=0$ (the BA 
scale-free network model).

\begin{figure}[!htb]
\begin{center}
\includegraphics[angle=-90,scale=0.55]{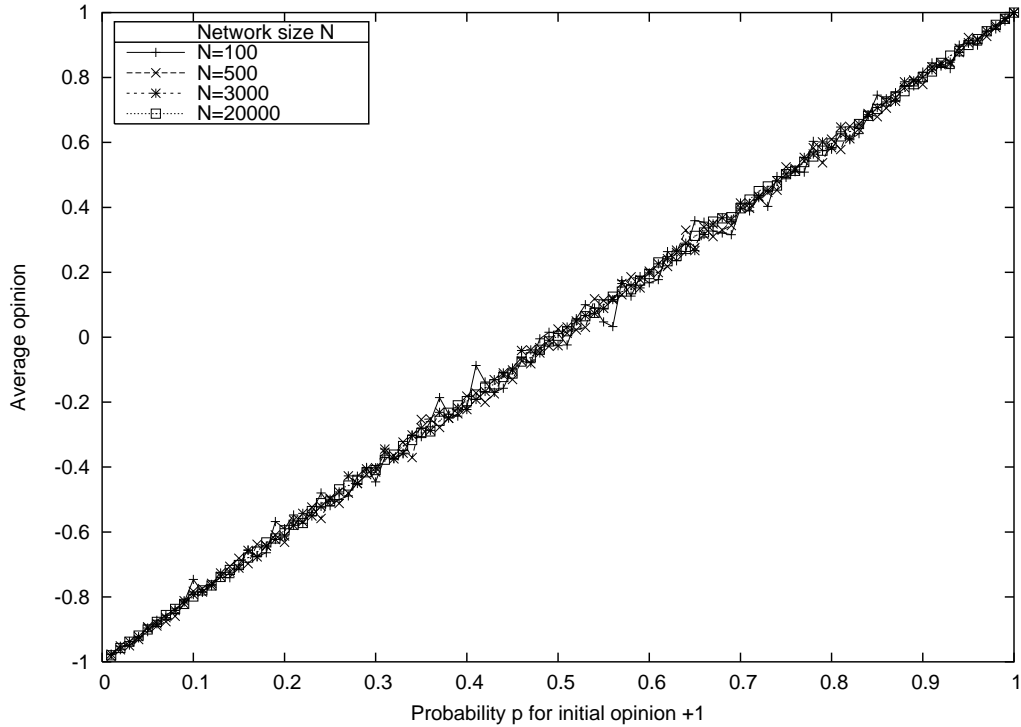}
\end{center}
\caption{The average opinion versus the probability $p$ for initial opinion
  $+1$, when $1$ node convincing strategy is considered for different 
network size $N$: $N = 100$ ($+$), $N = 500$ ($\times$), $N = 3000$ 
($\ast$) and $N = 20000$ ($\boxdot$).}
\label{fig:rule1}
\end{figure}

\subsection{The Sznajd model}
\subsection{Binary opinions}
Every node in the network is considered to be an individual with an opinion
that in the beginning of the simulation is randomly chosen with probability
$p$ for opinion $+1$. Once the network has been completely
constructed, we start the consensus process of Sznajd. At every time step
$t>0$, all the nodes are randomly visited and updated (a random list 
of nodes assures that each node is reached exactly once) by following 
four variations:

\begin{itemize}
\item {\bf 1 site convincing:} For each site $i$ chosen, we change the 
opinion of all its neighbors to the site's opinion.

\item {\bf 2 sites convincing:} For each site $i$ chosen, we select randomly 
one of its neighbors. If this selected neighbor has the same opinion as 
the site $i$, then all their neighbors follow the pair's opinion. Otherwise, 
nothing is done.

\begin{figure}[!h]
\begin{center}
\includegraphics[angle=-90,scale=0.55]{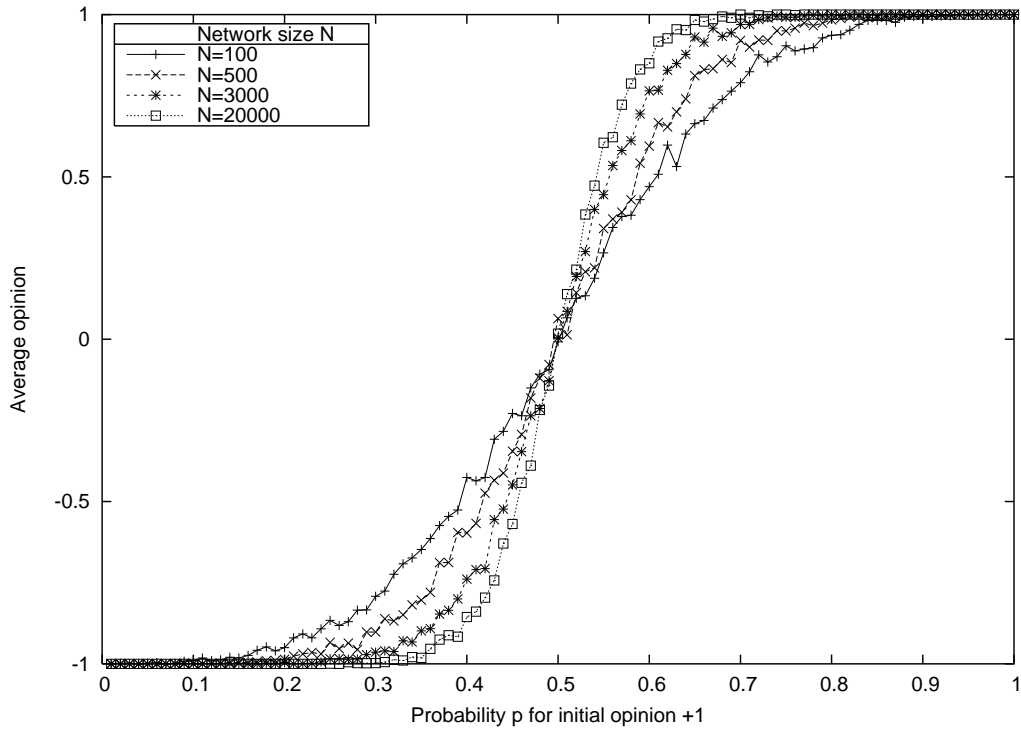}
\end{center}
\caption{As Fig. \ref{fig:rule1}, but for the $2$ nodes convincing case.}
\label{fig:rule2}
\end{figure}

\item {\bf 3 sites convincing:} For each $i$ site chosen, we select 2 of 
its neighbors at random. If all these three sites have the same opinion, 
they change the opinion of all their neighbors.

\item {\bf triad convincing:} For each $i$ site chosen, we select 2 of 
its neighbors at random. If all these three sites form a triangle and 
have the same opinion, they change the opinion of all their neighbors.
In the previous case, the 3 nodes must be connected with each other, 
they must not necessarily form a triangle.

\end{itemize}

\begin{figure}[!h]
\begin{center}
\includegraphics[angle=-90,scale=0.55]{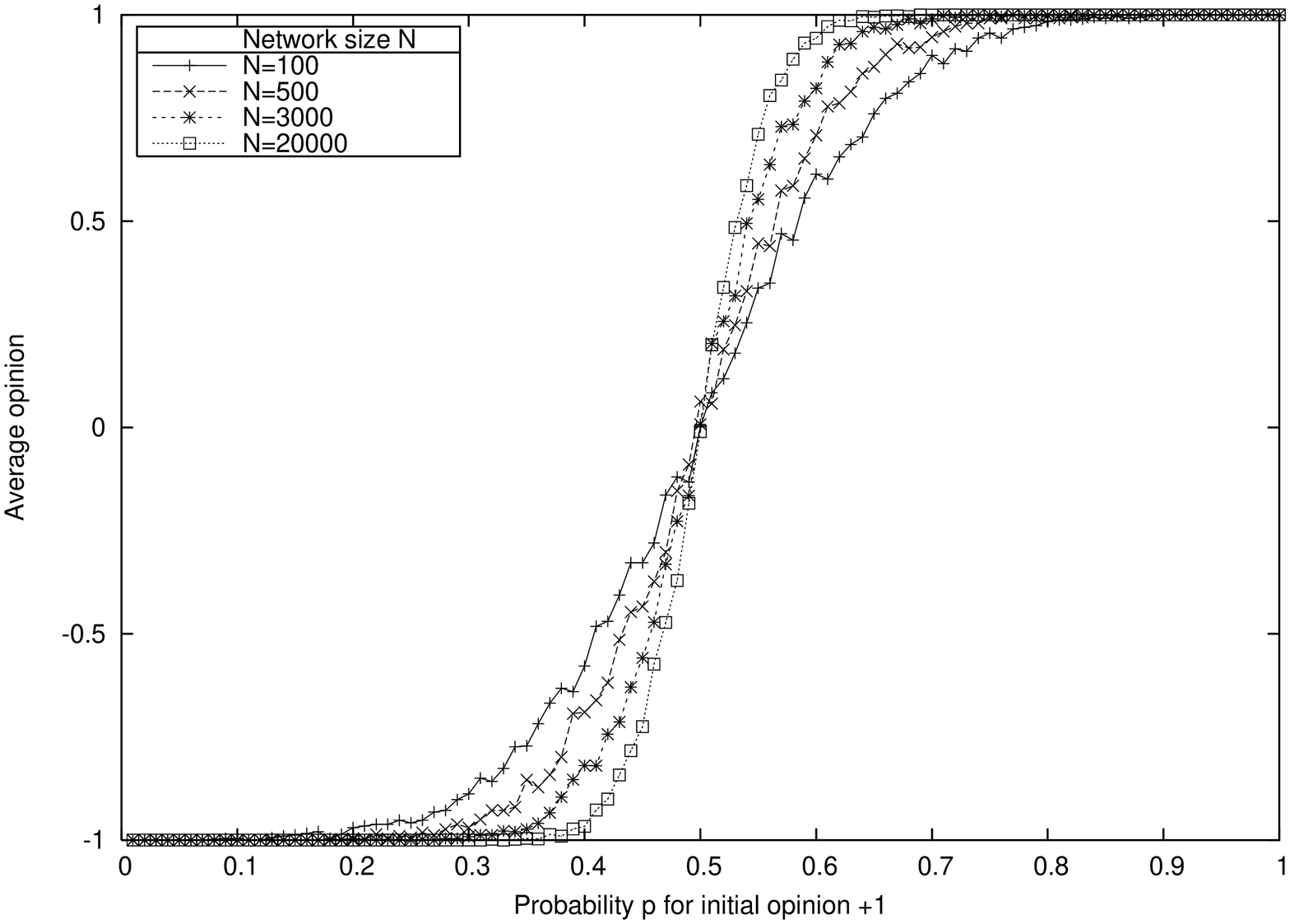}
\end{center}
\caption{As Fig. \ref{fig:rule1}, but for the $3$ nodes convincing case.}
\label{fig:rule3}
\end{figure}

In Figure (\ref{fig:rule_general}) we present the mean opinion of the 
network as a function of the initial concentration $p$ of opinion $+1$ 
for all the cases mentioned above after $t=1$ and $t=10$ timesteps. 
From this figure we can see that after $t=1$ timestep in all the cases 
a full consensus cannot be reached, however after $t=10$ timesteps the 
whole system always lead to a fixed point: consensus (all sites have 
opinion $+1$ or $-1$). Figures \ref{fig:rule1}, \ref{fig:rule2} and 
\ref{fig:rule3} correspond to the average of the results obtained for 
$1000$ samples ($1000$ different initial seeds for the random number 
generator). For 2 nodes, 3 nodes and triad (not shown) convincing rules 
one can observe a phase transition: concentration $p>1/2$ lead to full 
consensus $+1$ and concentrations $p<1/2$ to full consensus $-1$ 
for large enough systems. Since in a finite network, of course, phase
transitions are never sharp, and thus the transition is indicated 
numerically by a slope (Figs. \ref{fig:rule2} and \ref{fig:rule3}) 
becoming the steeper the larger the network size is. In an infinite 
network, one would get a sharp step function for the mean opinion of 
the sytems versus the initial concentration $p$ of opinion $+1$. This 
does not hold for the $1$ node convincing case (see Fig. \ref{fig:rule1}) 
where there is no such transition and the mean opinion of the system 
varies smoothly as a function of the initial density of opinion $+1$, 
as well the results does not have any dependence on the network size $N$. 
The phase transition at $p_{c}=1/2$ here observed does not exist in one
dimension \cite{sznajd1} or when a single site \cite{och} (instead of a pair or
plaquette) on the square lattice \cite{stauffer2} convinces its neighbors, 
although it is also found on the square lattice when a plaquette or a
neighboring pair persuades its neighbors \cite{stauffer2},  on a
correlated-diluted square lattice \cite{moreira}, on a triangular and simple
cubic lattice if a pair convinces its $8$ (or $10$, respectively) neighbors
\cite{chang}, on the Barab\'{a}si-Albert network \cite{bonnekoh} and on a 
pseudo-fractal network \cite{pseudo}.

\subsection{Multi-pluralist world}

Instead of beginning the simulation with only two opinions, $S_i=+1$ or 
$S_i=-1$, now we analyze the model when many different opinions 
are considered. After generating the triad network, as explained before, 
every individual assumes an opinion randomly selected from ${N_c}$ 
different opinions, with ${N_c}$ ranging from 2 opinions 
($S_i=+1$ or $S_i=-1$) to the network size $N$ ($S_i=1$ or $S_i=2$ 
or $S_i=3$ $\dots$ $S_i=N$) and $i=[1,2,3,\dots,N]$. At every time 
step $t>0$, all the nodes are randomly selected and updated by following 
the rules previously described.

\begin{figure}[!h]
\begin{center}
\includegraphics[angle=-90,scale=0.55]{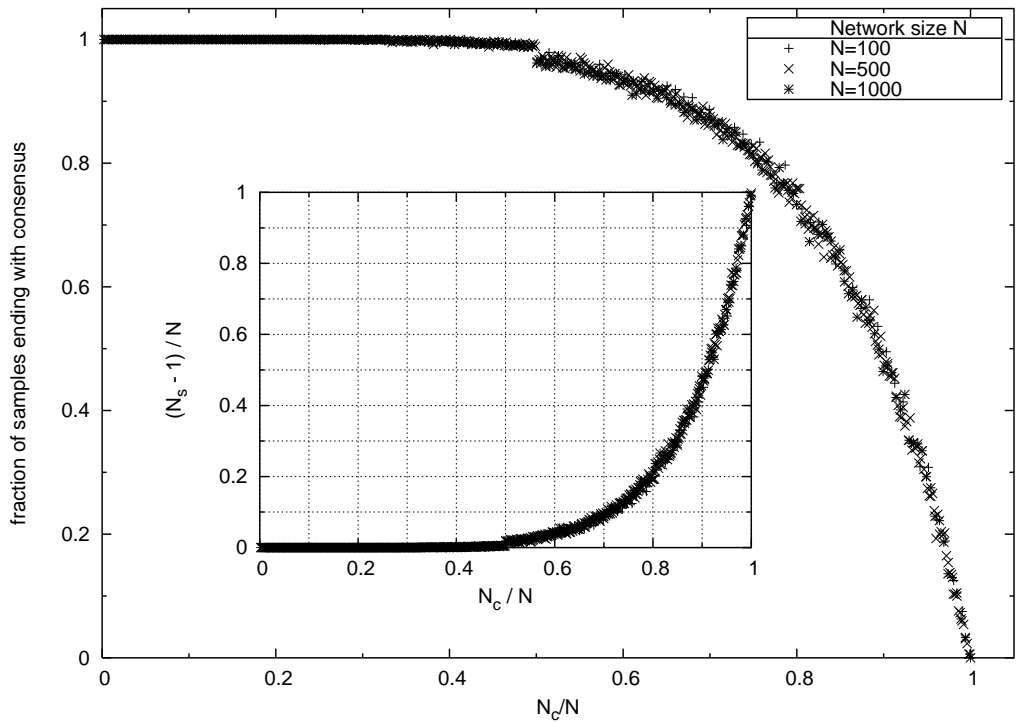}
\end{center}
\caption{Consensus fraction and scaled final number of different opinions 
($(N_{s} - 1)/ N$) versus the scaled number of different ones at 
the beginning of the simulation ($ {N_c} / N$) for the 2 nodes 
convincing rule and when all sites have a random initial opinion.}
\label{fig:pluri_rule2}
\end{figure}

\begin{figure}[hbt]
\begin{center}
\includegraphics[angle=-90,scale=0.55]{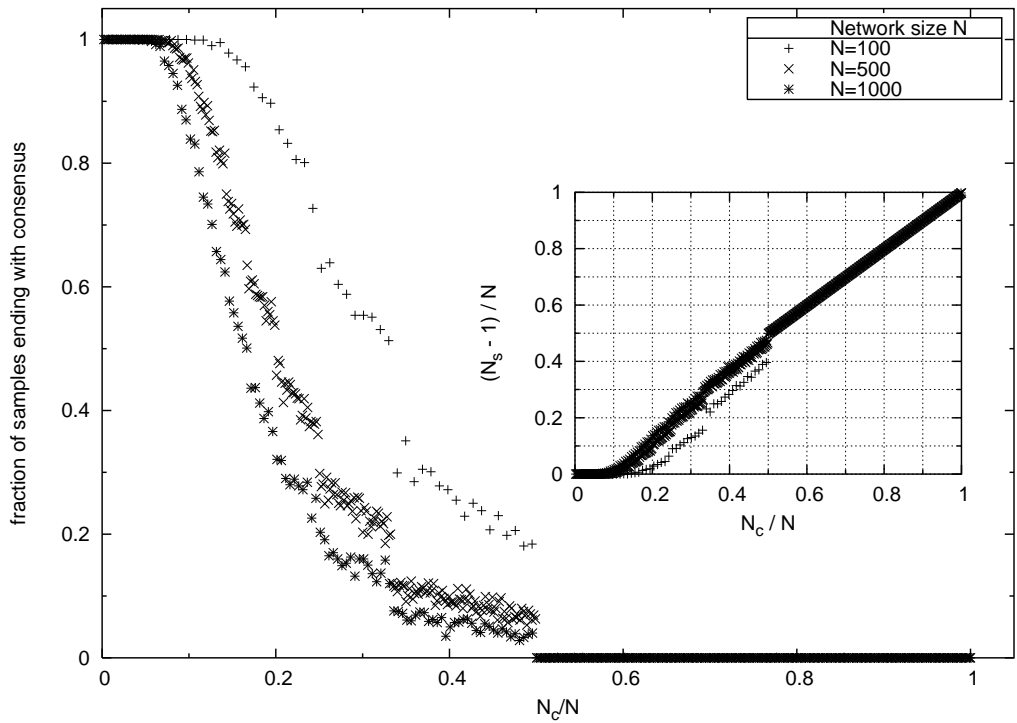}
\end{center}
\caption{As Fig. \ref{fig:pluri_rule2}, but for the 3 nodes convincing rule.}
\label{fig:pluri_rule3}
\end{figure}

In Figures \ref{fig:pluri_rule2} and \ref{fig:pluri_rule3} we present 
the fraction of samples, out of 1000, ending with full consensus 
as a function of the scaled number $N_{c}/N$ of different opinions 
at the beginning of the simulation. The insets show the scaled 
number $N_{s}$ of different opinions at the end of the 
simulation versus the scaled number $N_{c}$ of different ones at the 
beginning of the simulation. The $1$ node convincing rule (not shown) 
always leads the system to a full consensus (the whole system reaches 
a fixed point after some time-steps, i.e., all sites have the same 
opinion) independently of the initial number of different opinions 
and the network size $N$. Nevertheless, if the system evolves 
under the $2$ nodes convincing rule (Figure \ref{fig:pluri_rule2}), 
all the samples reach a fixed point with all the individuals having 
the same opinion ($N_{s}=1$, see inset) when ${N_c} < 0.50$; 
and for ${N_c} > 0.50$ the fraction of samples which has reached 
a full consensus decays slowly towards to zero as the scaled initial 
number ${N_c}$ of different opinions increases, as well the 
number $N_{s}$ of surviving opinions grows exponentially with 
${N_c}$ and independent of $N$. If the $3$ nodes or triad 
(not shown) convincing rule is considered (see Fig. \ref{fig:pluri_rule3}), 
only when ${N_c}<0.1$ the complete consensus ($N_{s}=1$, see inset) 
is found in all the samples, for $0.1 < {N_c} < 0.5$ the fraction
of samples decays slowly as ${N_c}$ increases and $N_{s}$ grows
to values closer to ${N_c}$, besides finite size effects appear; 
when ${N_c} > 0.5$ no sample reaches a consensus and $N_{s}$ 
roughly equals $N_{c}$, that means everybody keeps its own 
opinion and a fixed point is reached soon. 

Additionally, we have also simulated the $2$ nodes convincing rule 
when only a fraction ${N_c}$ of individuals at the beginning of the 
simulation assumes an opinion. In contrast to the same rule analyzed 
above where the simulation starts with all the individuals having an 
opinion, in such way if ${N_c} < {N}$ some individuals can assume 
the same opinion, now ${N-N_c}$ individuals do not assume any opinion 
($S_i=0$) and they are convinced to follow the neighbor's opinion according 
to probability $p_c$: a pair of neighbors in agreement convinces all 
the nodes connected with them with a probability for each of the two 
nodes inversely proportional to the time-independent number $k$ of 
nodes connected to it, $p_{c}=k^{-1}$. Alternatively, this case can be 
associated as an election campaign when ${N_c}$ candidates try to
persuade ${N}$ voters to vote for them. As in real elections, we do not
wait for a kind of equilibrium state, but count the votes at some 
intermediate time.

\begin{figure}[hbt]
\begin{center}
\includegraphics[angle=-90,scale=0.55]{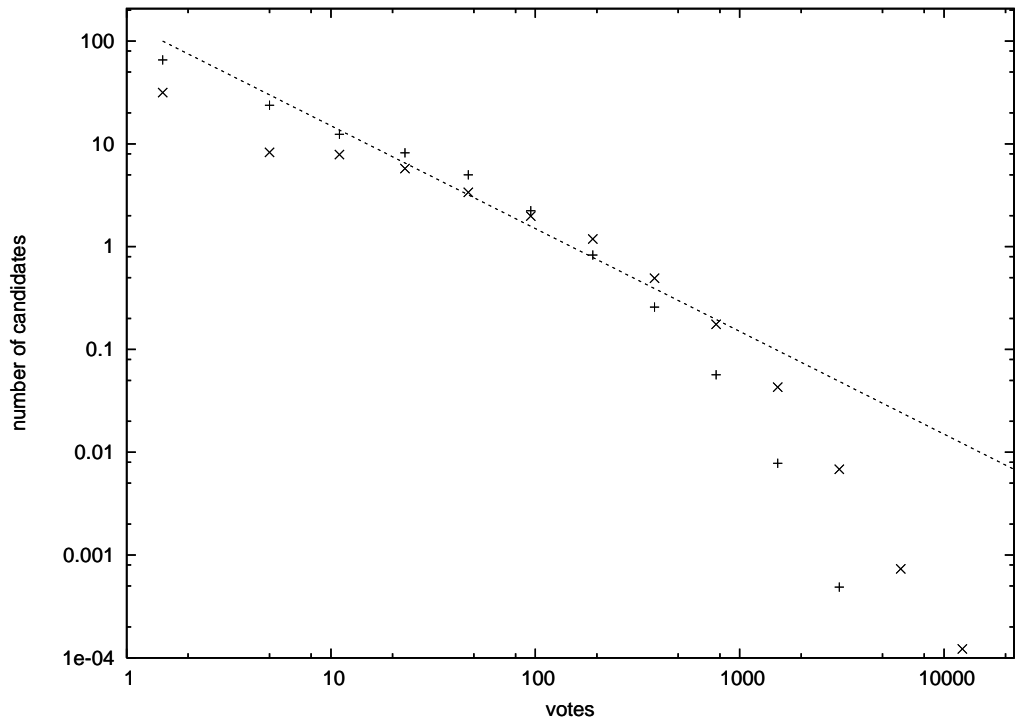}
\end{center}
\caption{Distribution of the number of candidates getting $v$ votes 
each at two different timesteps $t = 15$ ($+$) and $t = 20$ ($\times$).}
\label{fig:votes}
\end{figure}

Figure \ref{fig:votes} shows the distribution of the number of candidates 
getting $v$ votes each at two different timesteps $t = 15$ ($+$) and $t = 20$ 
($\times$). In agreement with findings for real elections which shows 
that the distribution of the number of candidates ${N_v}$ receiving 
a fraction of votes $v$ follows a power-law  ${N_v} \propto 1/v$ 
\cite{bernardes,pseudo,indo} except for downward deviations at small
and large numbers $v$ of votes, also our simulations obey this
hyperbolic law including the deviations at the ends.

\section{Conclusions}

By simulating the Sznajd model of socio-physics on a more realistic
network with stronger clustering, we have found a phase transition as a
function of the initial concentration of $S=+1$ opinions at $p_{c}=1/2$:
For $p < 1/2$ all samples end up with $S=-1$, and for $p > 1/2$  they all
end up in the other fixed point $S=+1$, for large enough networks. This
phase transition exists for $2$ nodes, $3$ nodes and triad convincing rule,
however it does not exist for $1$ node convincing rule. Instead of only two
opinions, but considering many opinions, we notice that $1$ node convincing
rule leads the system always to a complete consensus, while for the $2$ nodes 
convincing rule a full consensus is obtained when ${N_c}/N < 0.5$  and if 
${N_c}/N > 0.5$ the final number of different opinions grows exponentially 
with the number ${N_c}$ of initial ones; for $3$ nodes and triad convincing
rules a full consensus is observed only for ${N_c}/N < 0.1$, if 
$0.1 < {N_c} < 0.5$ the fraction of samples decays slowly as ${N_c}$ 
increases and $N_{s}$ grows to values closer to ${N_c}$ and when 
${N_c} > 0.5$ the final number $N_{s}$ of different opinions roughly 
equals $N_{c}$. When simple and minor changes are taken into account 
in the model, the hyperbolic law observed empirically in real
elections can be reproduced quite well.

\medskip
\noindent{\bf Acknowledgments}: The author thanks D. Stauffer for a critical 
reading of the manuscript; this work has been supported by a research grant 
from Alexander von Humboldt.

\end{document}